\documentclass{ieeeaccess}
\usepackage{cite}
\usepackage{amsmath,amssymb,amsfonts}
\usepackage{algorithmic}
\usepackage{graphicx}
\usepackage{textcomp}

\usepackage{listings}
\usepackage{hyperref}

\def\BibTeX{{\rm B\kern-.05em{\sc i\kern-.025em b}\kern-.08em
    T\kern-.1667em\lower.7ex\hbox{E}\kern-.125emX}}
    
\graphicspath{{./images/}}
\begin{document}
\history{Received March 19, 2022, accepted April 13, 2022, date of publication April 22, 2022, date of current version May 2, 2022.}
\doi{10.1109/ACCESS.2022.3169744}

\title{Symbolic music generation conditioned on continuous-valued emotions}
\author{\uppercase{Serkan Sulun}\authorrefmark{1},
\uppercase{Matthew ~E.~P.~Davies\authorrefmark{2}}
, and \uppercase{Paula Viana}\authorrefmark{1,3}
\IEEEmembership{Senior Member, IEEE}}
\address[1]{Institute for Systems and Computer Engineering, Technology and Science (INESC TEC), 4200-465 Porto, Portugal}
\address[2]{University of Coimbra, Centre for Informatics and Systems of the University of Coimbra, Department of Informatics Engineering, 3030-290 Coimbra, Portugal}
\address[3]{Polytechnic of Porto, School of Engineering, 4200-072 Porto, Portugal}
\tfootnote{The work of Serkan Sulun was supported by the ‘‘la Caixa’’ Foundation under Grant 100010434 and Grant LCF/BQ/DI19/11730032. The
work of Matthew E. P. Davies was supported by FCT—Foundation for Science and Technology, I.P., through the Project MERGE through
the National Funds (PIDDAC) through the Portuguese State Budget under Grant PTDC/CCI-COM/3171/2021; and in part by the European
Social Fund through the Regional Operational Program Centro 2020 Project CISUC under Grant UID/CEC/00326/2020.}

\markboth
{Sulun \headeretal: Symbolic music generation conditioned on continuous-valued emotions}
{Sulun \headeretal: Symbolic music generation conditioned on continuous-valued emotions}

\corresp{Corresponding author: Serkan Sulun (e-mail: serkan.sulun@inesctec.pt).}

\begin{abstract}
In this paper we present a new approach for the generation of multi-instrument symbolic music driven by musical emotion. 
The principal novelty of our approach centres on conditioning a state-of-the-art transformer based on continuous-valued valence and arousal labels. In addition, we provide 
a new large-scale dataset of symbolic music paired with emotion labels in terms of valence and arousal.
We evaluate our approach in a quantitative manner in two ways, first by measuring its note prediction accuracy, and second via a regression task in the valence-arousal plane.
Our results demonstrate that our proposed approaches outperform conditioning using control tokens which is representative of the current state of the art. 
\end{abstract}


\titlepgskip=-15pt

\maketitle

\section{Introduction}
\label{sec:introduction}

\textit{Affective algorithmic composition} (AAC) deals with automatic composition of music based on specific emotions \cite{aac}. The use cases of AAC include composing soundtracks for videos and video games \cite{soundtrack}, neurofeedback training for medical use \cite{medicinal} and developing brain-computer music interfacing systems \cite{bci}. Although the relationship between music and emotion is well-studied \cite{music_and_emotion}, choosing the ``optimal'' emotion model to investigate this relationship is still a debated subject \cite{emotion_models_for_music}. Existing work on AAC mostly uses approaches that belong to two main categories of emotion models, namely categorical and dimensional \cite{aac}. Categorical emotion models use discrete labels such as happy, sad, angry, and surprised \cite{discrete_emotions}. The studies on dimensional approaches argue that categorical approaches are insufficient for modeling the complexities and subtleties of human emotions, and propose using continuous-valued coordinates that locate points on a low-dimensional space \cite{emotion_survey}. 

The most common dimensional emotion model is Russell's \textit{circumplex model of affect}, which is a two-dimensional model consisting of valence (unpleasantness vs. pleasantness) and arousal (relaxed vs. aroused) dimensions \cite{russell}. Russell also maps categorical emotions onto this valence-arousal plane, providing its exemplary usages, such as the categorical emotion ``calm'' having high valence and low arousal values, and ``annoyed'' having low valence and high arousal values. 

The early works on AAC used various melodic, harmonic, and rhythmic features to target specific emotions (see the overview in \cite{aac}). However, in these works, the correspondence between emotions and musical features was only approximate, and it was generally established using discrete features and categorical emotions \cite{features_emotion_hevner, features_emotion_levi}. The advent of deep learning enabled using complex models on large labeled datasets and eliminated the necessity of using intermediate features \cite{alexnet}. Recent AAC models are trained on datasets containing symbolic music and emotion labels, in an end-to-end fashion \cite{vgmidi,emo_lstm,emopia}. However, these works could only use a very small number of categorical emotion labels, possibly due to the small sizes of their training datasets.

In this work, we introduce an AAC model that can be conditioned on continuous coordinates on the valence-arousal plane. This approach enables the representation of complex emotions with their subtleties. To this end, we combine several datasets, resulting in a labeled MIDI dataset two orders of magnitude greater than existing labeled MIDI datasets. Using this dataset, we successfully train large transformer models \cite{transformer} on a single GPU, to generate multi-instrument symbolic music conditioned on emotion. To the best of our knowledge, this is the first music generation model that can be conditioned on both valence and arousal simultaneously, therefore enabling conditioning on an arbitrary emotion from the widely-used circumplex model of affect. The main contributions of our work are as follows:

\begin{itemize}
    \item We create a symbolic music dataset with continuous-valued labels. These labels can be mapped onto the valence-arousal plane, bridging symbolic music and perceived emotion. Although this dataset has weak labels, it is two orders of magnitude larger than the existing datasets \cite{vgmidi, mirexlike, emopia}.
    \item We propose multiple architectures for conditional symbolic music generation which outperform the state-of-the-art architecture in quantitative evaluation.
    \item Our proposed models additionally allow the usage of continuous condition values, with the capability of dynamic conditioning in which a user could arbitrarily change the condition values throughout the generation.
\end{itemize}

The remainder of the paper is structured as follows. In Section \ref{sec:related}, we discuss the existing work on sequence modeling, symbolic music and musical emotion datasets. In Section \ref{sec:methodology}, we explain pipeline of dataset creation, model implementation, training and inference. In Section \ref{sec:eval} we mention methods of evaluation. Finally in Section \ref{sec:results} we present and discuss the quantitative results.

\section{Related work}
\label{sec:related}

\subsection{Generic sequence modeling}

Symbolic music can be represented as sequential data, similar to text. Hence, the same models can, in principle, be used for both natural language processing (NLP) and symbolic music processing. One of the oldest neural network architectures for sequence modeling is the recurrent neural network (RNN), where a single input sample (token) is processed at each timestep. The network is trained by calculating the gradient of the error across each timestep, using the algorithm named \textit{backpropogation through time}. However, RNNs aren't very successful in modeling long sequences, because as the sequence grows longer, the backpropagated gradients can approach zero. This problem is named \textit{vanishing gradient problem}. Long short-term memory (LSTM) networks alleviate this problem by using specialized gates \cite{lstm}. A similar flavor of RNN named gated recurrent unit (GRU) can achieve similar performance to LSTM, using a simpler architecture with fewer parameters \cite{gru}. But even these new flavors of RNN aren't efficient in processing long sequences, and they tend to ``forget'' the old input samples as the sequence grows longer. The attention mechanism addresses this problem by explicitly modeling the dependencies between all pairs of input samples \cite{attention}. Finally, the transformer model achieved the current state-of-the-art results in sequence processing by incorporating the attention mechanism in a multi-headed and multi-layered architecture \cite{transformer}. 

The original transformer implementation had an encoder and a decoder network, and it was tested on the task of machine translation. It is common to use encoder-decoder architectures for machine translation, where the encoder processes the source text and the decoder generates the output \cite{encoder_decoder_rnn}. Since the task of language modeling involves generating text from scratch, it can be seen as analogous to music generation, hence both tasks can be categorized under the task of \textit{sequence generation}. Because there are no separate source and target sequences, state-of-the-art language models only consist of a decoder \cite{gpt3}. Sequence-generating neural networks are trained with input and target sequences, which belong to the same domain. Specifically, the target sequence is one timestep shifted version of the input sequence, hence for each input token, the network predicts the next token.

\subsection{Conditional natural language processing}

Although it is a loosely used term, \textit{conditioning} refers to controlling a model's output by providing auxiliary inputs, i.e., conditions. The conditions can belong to the same domain as the input and the target, so it can be possible to train the model using unlabeled data. Alternatively, they can also belong to different domains, such as the labels of a labeled dataset. Even the earliest neural networks for natural language processing made use of conditioning. Mikolov and Zweig developed a language model using conditions such as topic and genre, where a conditioning vector was created using a linear layer and concatenated with the hidden state of the recurrent neural network (RNN) \cite{conditionalLM}. Sennrich et al. developed an encoder-decoder RNN model for translation from English to German, conditioned on politeness \cite{conditional_translator}. To achieve this they used control tokens specifying the user's preference for a formal or informal translation. Conditional Transformer Language (CTRL) model feeds control tokens which denote domain, style, topics, etc., into a large transformer, obtaining state-of-the-art results in conditional language modeling \cite{ctrl}. Krishna et al. performed style transfer by generating paraphrases, and showed that training separate models for each style outperforms training a single model that uses style-specific control tokens \cite{paraphrase}. Sheng et al. identified triggers, i.e. subsequences that generate biased text when fed as inputs, and use them as primers to induce or balance bias in language modeling \cite{bias}. Smith et al. \cite{dialogue} investigated controlling the style of dialogue generation, by comparing three methods, namely, retrieve-and-refine \cite{retrieve}, inference-time iterative refinement \cite{iterative} and conditional generation using control tokens \cite{ctrl}. They showed that conditional generation using control tokens outperforms other methods.

While the majority of the works in the literature use categorical variables, such as control tokens, to control language modeling, the problem of image captioning can be formulated as a text generation task based on images, which are non-categorical variables. 
Here, the input image is usually processed with a convolutional neural network, and the resulting features are used for conditioning a separate language model. While earlier works used RNN as the language model \cite{image_caption_rnn}, state-of-the-art models replaced it with a transformer \cite{image_caption_transformer}. Zhu et al. compared different conditioning methods and observed similar performances \cite{image_caption_transformer}. These methods include, feeding the spatial image features into the cross-attention layer of the decoder \cite{image_caption_attention}, combining image feature with each word embedding, and feeding the image feature before the word embeddings \cite{image_caption_rnn}. 

\subsection{Symbolic music generation}

State-of-the-art symbolic music generators make use of large unlabeled symbolic music datasets. The Lakh MIDI dataset \cite{lakh} is a collection of $176581$ unlabeled multi-instrument MIDI files, $45129$ of which have been matched to $31034$ entries in the Million Song Dataset \cite{msd}. To the best of our knowledge, there are only three publicly available symbolic music datasets with emotion labels, although their sample sizes are very small. VGMIDI consists of $204$ video game soundtracks played by piano and has continuous-valued labels for valence and arousal \cite{vgmidi}. Panda et al. have created a music dataset with discrete emotion labels. The dataset mostly contains audio files and emotion labels, but for $193$ samples, the MIDI files are available \cite{mirexlike}. The EMOPIA dataset contains clips extracted from $387$ songs and annotated using discrete labels corresponding to the four quadrants of the two-dimensional circumplex model of affect \cite{emopia}.

Early works employing neural networks for music generation used recurrent neural networks \cite{blues_lstm}. However, the recent advent of the transformer model has enabled the usage of much longer dependencies. The music transformer built upon the transformer model by incorporating relative positional information, obtaining state-of-the-art results in symbolic music generation \cite{musictransformer}. 

It should be noted that any sequence generator can be conditioned using a sub-sequence as a \textit{primer} at inference time. In symbolic music generation, this corresponds to feeding some melody to the model and predicting the melody that follows it. In this method, the condition and the target belong to the same domain, hence the models can be trained using unlabelled data. Other symbolic music generation tasks that utilize same-domain conditioning are accompaniment generation \cite{coconet, musegan}, interpolation \cite{musicvae}, inpainting \cite{sketchnet, music_machine}, and style transfer \cite{musical_style, wang}. MidiNet can generate melodies that are conditioned on chords, by training on a private dataset that includes chord information \cite{midinet}. OpenAI's MuseNet model, the state-of-the-art, is trained on a combination of datasets, and the generation can be conditioned on specific artist names, genres, or styles, using primer control tokens \cite{musenet}.

It is also possible to use low-level symbolic music features for conditioning \cite{tonal_tension, pati, yang}. These features such as tempo, note density, pitch range, and tonal tension, can be calculated automatically, hence there is no need for a labeled dataset.  Tan and Herremans aimed at compensating for the small size of the labeled VGMIDI dataset, hence they augmented it with the unlabeled MAESTRO (MIDI and Audio Edited for Synchronous TRacks and Organization) dataset \cite{maestro} using low-level rhythm and note density features to infer the high-level arousal feature \cite{fadernets}.

The creators of the VGMIDI dataset also devised a method for symbolic music generation conditioned on emotion \cite{vgmidi}. Using a genetic algorithm, they fine-tuned the weights of a pretrained LSTM. This was done separately for positive and negative valence conditions, resulting in two models. Both Zhao et al. and Hung et al. generated symbolic music conditioned on four categorical emotions belonging to the four quadrants of the valence-arousal plane \cite{emo_lstm, emopia}. Zhao et al. \cite{emo_lstm} labeled the piano-midi dataset \cite{pianomidi} using categorical labels, and trained a biaxial LSTM \cite{balstm} on this labeled dataset. Hung et al. \cite{emopia}, the creators of the EMOPIA dataset, trained a transformer model that is conditioned using control tokens \cite{ctrl}.

\subsection{Spotify audio features}

The \textit{Spotify for Developers} application programming interface (API) allows users to access audio features for a given song from Spotify's private database \cite{spotify}. These audio features are both low- and high-level and are namely danceability, energy, key, loudness, mode, speechiness, acousticness, instrumentalness, liveness, valence, and tempo. The high-level features such as valence are estimated using machine learning algorithms that are trained on data labeled by experts \cite{spotify_echonest, echonest_valence}.

\section{Methodology}
\label{sec:methodology}

\subsection{Lakh-Spotify dataset}
\label{sec:dataset}

To create a dataset that contains pairs of MIDI files and high-level labels, we use the Spotify for Developers API and obtain audio features for the samples from the Lakh MIDI dataset (LMD). In particular, we use the \textit{LMD-matched} subset, since its samples are matched to the entries in the Million Song Dataset (MSD) \cite{msd}, hence we can use the metadata from MSD to search Spotify's database. Using the track ID for each MIDI file, we first obtain the song title, artist name, and Echo Nest song ID. Using the Echo Nest song IDs, and another dataset named \textit{Million Song Dataset Echo Nest mapping archive} \cite{mapping}, we also obtained Spotify track IDs. 

Next, for each MIDI sample, we conducted a search using the Spotify for Developers API. The query for the search was the associated Spotify ID. If the Spotify ID was not available, we used the artist name and the song title as the query. 
The entire dataset creation pipeline can be seen in Figure \ref{fig:data}. 

\Figure[h]()[width=0.99\linewidth]{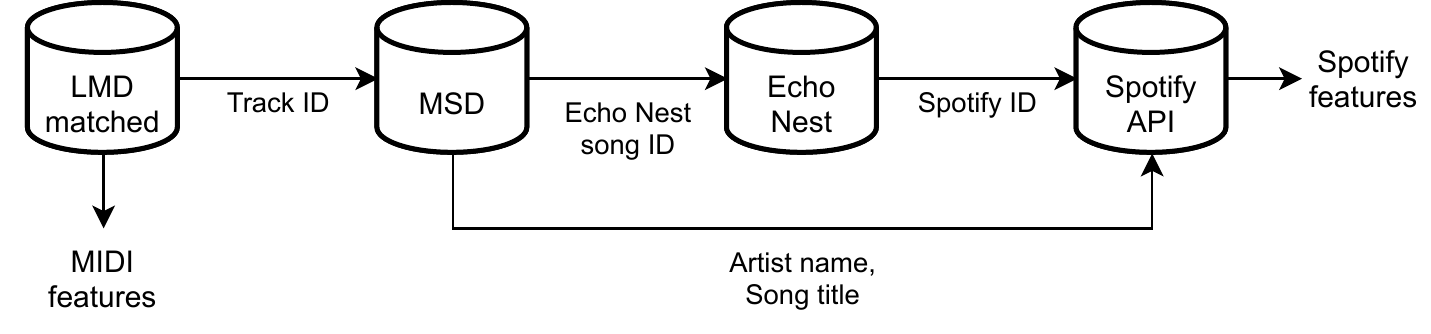}
{Dataset creation pipeline. The MIDI features are note density, estimated tempo, and the Spotify features are danceability, energy, key, loudness, mode, speechiness, acousticness, instrumentalness, liveness, valence, and tempo. \label{fig:data}}

\begin{table}[]
\caption{Our dataset compared to other MIDI dataset with emotion labels.}
\label{table:data comparison}
\centering
\begin{tabular}{|c|c|c|c|}
\hline
Dataset                                                           & \begin{tabular}[c]{@{}c@{}}Number of \\ songs\end{tabular} & Average duration & Label type                                                          \\ \hline
\hline
Panda et al. \cite{mirexlike}                                                     & 193                                                         & 216.1 s.         & \begin{tabular}[c]{@{}c@{}}Categorical \\ (discrete)\end{tabular}   \\ \hline
\begin{tabular}[c]{@{}c@{}}Ferreira and \\ Whitehead \cite{vgmidi}\end{tabular} & 204                                                         & 112.4 s.         & \begin{tabular}[c]{@{}c@{}}Dimensional \\ (continuous)\end{tabular} \\ \hline
Hung et al. \cite{emopia}                                                      & 387                                                         & 102.3 s.         & \begin{tabular}[c]{@{}c@{}}Categorical \\ (discrete)\end{tabular}   \\ \hline
Ours                                                              & 34791                                                       & 223.7 s.         & \begin{tabular}[c]{@{}c@{}}Dimensional \\ (continuous)\end{tabular} \\ \hline
\end{tabular}
\end{table}

Because the Spotify features belong to the audio versions of each track, they can only be considered ``weak'' labels for the MIDI versions. Thus, to improve our dataset, we also included the low-level MIDI features such as note density, i.e., number of notes per second, the estimated tempo, and the number of instruments. These low-level features can also be used to model the arousal dimension of the circumplex model of affect \cite{aac, fadernets}.

For completeness, we derived the low-level MIDI features for the entire Lakh MIDI dataset, labeled as \textit{LMD-full}, where not all the samples are necessarily mapped to the entries in the MSD. The LMD-full dataset consists of $178561$ MIDI files. Upon investigation, we found that $174270$ of those are valid, and we discarded the remaining corrupt or empty files. The Lakh MIDI dataset was constructed by downloading MIDI files from publicly-available sources on the internet and then keeping the unique files according to their hash values. But upon examination, we saw that MIDI files with different hash values could still have the same musical content, possibly due to the difference in their metadata. To further filter the data to keep the MIDI files with unique musical content, we converted the MIDI files to piano rolls, using the pretty\_midi packages, and then re-calculated the hash values. As a result, we ended up with $152968$ MIDI files with unique musical content.

The matched split of the Lakh MIDI dataset, namely \textit{LMD-matched}, consists of $31034$ tracks from the MSD matched with $116189$ MIDI files from the LMD. Multiple MIDI files can be matched to the same track, and multiple tracks can be matched to the same MIDI file. Since our overall aim is to create a MIDI dataset with labels, we only kept MIDI files with unique musical content as we have done for the \textit{LMD-full} data split. Furthermore, we only kept the best matching track from the MSD for each MIDI file, based on the matching scores, in order to have only one set of labels for each MIDI file. As we have also done with \textit{LMD-full}, after keeping valid MIDI files with unique musical content, we ended up with $36545$ MIDI files that are matched with entries from the MSD. Based on the metadata from the MSD, we searched Spotify's dataset and were able to obtain audio features for $34791$ MIDI files.

\hspace*{-\parindent}
\begin{minipage}[t]{\linewidth}
\begin{lstlisting}[frame=single,basicstyle=\small,caption=A sample entry from proposed dataset.\vspace{4mm},label=dataset]
"cc992d0d8e82d09b7fe2466cf851497a": {

"midi_features": {
    "note_density": 30.364985431879415,
    "tempo": 84.000084000084,
    "n_instruments": 10
},
"matched_features": {
    "track_id": "TRUHHPK12903CBA84F",
    "match_score": 0.7362919446232072,
    "song_id": "SOSYWZT12AB0187E45",
    "title": "In The Summertime",
    "artist": "Mungo Jerry",
    "release": "Uber 30 - das rockt!",
    "spotify_id": "5VPOrzHyuULaiCKnwQNNCN",
    "spotify_title": "In The Summertime",
    "spotify_artist": "Mungo Jerry",
    "spotify_album": "Uber 30 - das rockt!",
    "spotify_audio_features": {
        "danceability": 0.681,
        "energy": 0.509,
        "key": 4,
        "loudness": -8.504,
        "mode": 1,
        "speechiness": 0.0461,
        "acousticness": 0.497,
        "instrumentalness": 2.72e-06,
        "liveness": 0.188,
        "valence": 0.963,
        "tempo": 82.614,
        "type": "audio_features",
        "id": "5VPOrzHyuULaiCKnwQNNCN",
        "uri": 
"spotify:track:5VPOrzHyuULaiCKnwQNNCN",
        "track_href": "https://api.spotify.c
om/v1/tracks/5VPOrzHyuULaiCKnwQNNCN",
        "analysis_url":"https://api.spotify.c
om/v1/audio-analysis/5VPOrzHyuULaiCKnwQNNCN",
        "duration_ms": 210387,
        "time_signature": 4
    }
}
}
\end{lstlisting}
\end{minipage}

In its complete form, we created a dataset which we name the \textit{Lakh-Spotify dataset}, that is supplementary to LMD-matched dataset. We show a sample entry and the included features in Listing \ref{dataset}. While the precise implementation details for their retrieval are not publicly available, an explanation of the Spotify audio features can be found in the online documentation\footnote{\url{https://developer.spotify.com/documentation/web-api/reference/}}. A comparison between our dataset and existing MIDI datasets with emotion labels is shown in Table \ref{table:data comparison}.

\subsection{Emotion-based music generation}
\subsubsection{Training data and pre-processing}

For our music generation task, we first pre-train our non-conditional vanilla model on the Lakh Pianoroll Dataset (LPD) \cite{musegan}, specifically the LPD-5-full subset. This subset is created by merging the individual tracks in the MIDI files into five common categories: drums; piano; guitar; bass; and strings. We chose to use this dataset to have a finite number of tokens since we represent the instruments explicitly, using separate note-on and note-off tokens for each instrument, similar to Payne et al. \cite{musenet} and Donahue et al. \cite{lakhnes}. After pre-processing, the non-conditional training data split has $96119$ songs. 

To train our conditional models, we first transfer the available weights from the vanilla model and then fine-tune on the LPD-5-matched dataset, namely the 5-instrument piano roll counterpart of the LMD-matched dataset. Since we previously generated low- and high-level labels for this dataset as explained in Section \ref{sec:dataset}, we used these labels for conditioning. After pre-processing, the conditional training data split has $27361$ songs. 

Before tokenization, we convert the piano rolls into MIDI, using the Pypianoroll package \cite{pypianoroll}. We use the pretty\_midi package for processing the MIDI data \cite{pretty_midi}. For tokenization, we use the event-based MIDI representation \cite{event}. During pre-processing, we filter out MIDI note-on and note-off events that have a pitch outside the range of the piano, i.e., lower than $21$ and higher than $108$, since these notes aren't audible using standard MIDI soundfonts. We use $125$ time shift tokens spanning the range from $8$~ms. up to $1$~s., in increments of $8$~ms., as done by Oore et al. \cite{event}. Adding a \texttt{<START>} token denoting the beginning of a sequence, and a \texttt{<PAD>} token to pad the sequences when necessary, we end up with $1007$ tokens for our vanilla (non-conditional) model.

\Figure[!t]()[width=\textwidth]{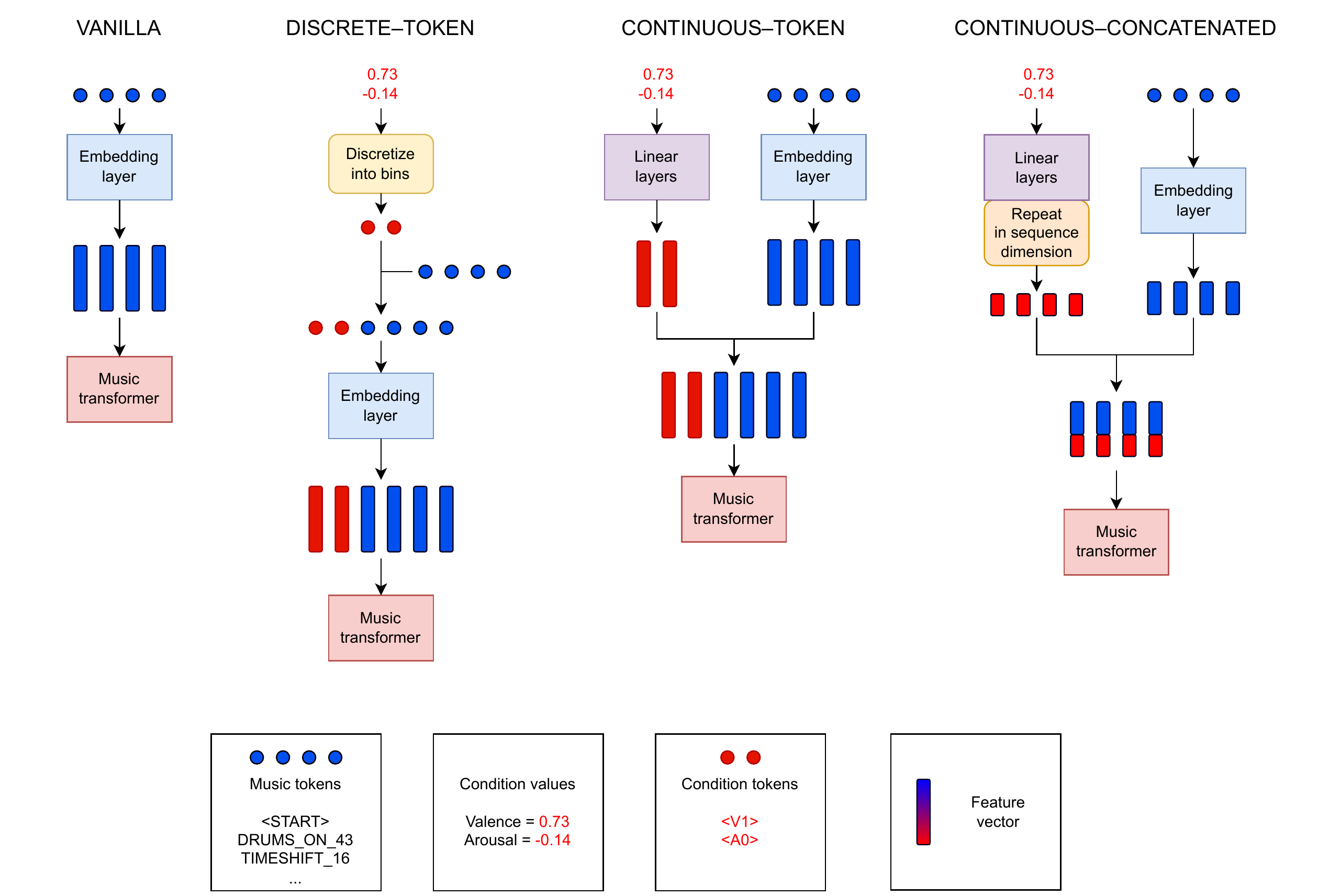}
{Models. \label{fig:models}}

The training input sequences are fixed-sized chunks, extracted from the MIDI sequences. With a probability of $0.5$ the beginning of a chunk corresponds to the beginning of a random bar, and the \texttt{<START>} token is inserted at the beginning. Otherwise, the chunk is extracted from a completely random location, and \texttt{<START>} token is not used. We found that this process is necessary to be able to generate sequences that are longer than the training input length, during inference. We transpose the pitches of all instruments except drums, by a randomly chosen integer value between $-3$ and $3$ inclusive. We chose this relatively narrow range of values to avoid any possible altering of the emotional character of the songs. Using these two methods to create the inputs, the effective training data size becomes much larger than the number of songs in the training data split.

We use two conditioning values to model valence and arousal, the valence feature from Spotify's database and the MIDI note density averaged over the number instruments respectively. Our preliminary experiments showed that using the Spotify audio feature for valence for conditional generation did yield meaningful output, but the energy feature wasn't as useful as the arousal feature. This is because both energy and arousal features are very sensitive to timbre, and the MIDI format is inadequate for representing timbre. As a solution, we used the note density of the MIDI files as the conditioning arousal feature for music generation, as suggested by Williams et al. \cite{aac}, and obtained satisfactory results in terms of musical coherence, similar to Tan and Herremans \cite{fadernets}. In its original form, Spotify's valence feature takes values between $0$ and $1$. For consistency with the two-dimensional circumplex model of affect and normalization purposes, we shifted and scaled both conditioning elements to take values between $-1$ and $1$. 

To create the testing data split, we ordered the file names from the LPD-matched subset alphabetically, and reserved the last 5\%, resulting in $1437$ songs. We stress that this testing data is not used during the non-conditional pre-training or conditional fine-tuning. 

When pre-processing the entire dataset, we filtered out the songs that contain fewer than $3$ instrument tracks. We then removed samples that constitute outliers considering their valence and arousal values. The threshold values for outliers were found by multiplying the interquartile range by $1.5$, then subtracting this value from the first quartile and adding this value to the third quartile. Concerning the distribution of valence values, we noticed a large peak located exactly at $-1.0$ (corresponding to $0.0$ in its original form), possibly due to invalid values, and hence we also removed those samples.

\subsubsection{Models}
\label{sec:models}

The backbone of our models is the music transformer \cite{musictransformer}, which is a decoder-only transformer using relative position embeddings. It has $20$ layers and a feature dimension of $768$. Each layer has $16$ heads and a feed-forward layer with a dimension of $3072$. Overall, our model has around $145$ million parameters. 

We experimented with different methods for conditioning the music generation process on the emotion features. We first implemented the current state-of-the-art approach in conditional sequence generation \cite{ctrl, musenet, emopia}, which we name \textit{discrete-token}, where we put the valence and arousal values into discrete bins and then converted them into control tokens. In detail, we quantize the condition values using $5$ equal-sized bins, where the central bin index is $0$. We chose the number of bins to model typical verbal quantifiers \textit{very low}, \textit{low}, \textit{moderate}, \textit{high} and \textit{very high}. The control tokens belonging to valence and arousal are placed before the music tokens, i.e., concatenated in the sequence dimension, only if the sample was taken from the beginning of a certain bar. The resulting sequence is then fed into the transformer. One of the disadvantages of this model is that, during inference, after the generation length reaches the input length, the inputs are truncated from the beginning, hence the control tokens are not fed. Another disadvantage is the information loss due to the binning of continuous values.

In our next approach, named \textit{continuous-token}, we use the normalized condition values in their continuous form. We feed each value to a separate linear layer, creating condition vectors that have the same length as the music token embeddings. Next, the condition vectors and music token embeddings are concatenated in the sequence dimension and fed into the transformer. During inference, and even after the generation length reaches the input length, we still insert the condition vectors at the beginning of the input sequence.

Our final approach is named \textit{continuous-concatenated}, where we create a single vector for the two normalized continuous condition values, repeat this vector in the sequence dimension, and concatenate it with every music token embedding. The lengths of the conditioning vectors and token embeddings are $192$ and $576$ respectively so that the total feature length of the transformer input is constant across models. All conditional models are trained by fine-tuning the pretrained vanilla (non-conditional) model. The representations of the models can be seen in Figure \ref{fig:models}.

\subsection{Implementation details}
We implemented our models using the Pytorch library \cite{pytorch} and trained them on a single NVIDIA Quadro RTX 6000 GPU. We used the Adam optimizer \cite{adam} with a learning rate of $2e{-5}$. Our preliminary experiments confirmed the findings of Donahue et al. \cite{lakhnes}, that common learning rates for language modeling tasks, about $2e{-4}$, are too high for MIDI generation tasks. We reduced the learning rate to $2e{-6}$ when the training loss plateaued and kept training until convergence. We used gradient clipping at a norm of $1$, with a dropout rate of $0.1$, a batch size of $4$, and an input length of $1216$. We used an autoregressive mask to prevent the model from attending to future tokens.

\subsection{Inference}
At inference, and before the generation starts, the input sequence only consists of the \texttt{<START>} token, except for the \textit{discrete-token} model where we also prepend two condition tokens for valence and arousal. For the models \textit{continuous-token} and \textit{continuous-concatenated}, the condition values are fed in parallel at every timestep, as explained in Section \ref{sec:models}. We generate the output autoregressively, where the generated token is appended to the input sequence, forming the input sequence for the next timestep. When the maximum input length of $1216$ is reached, we use the last $1216$ tokens of the generated sequence as the input, so that the maximum input length is not exceeded. We use nucleus sampling with $p=0.7$ from the temperature adjusted softmax distribution \cite{nucleus}, with a temperature of $1.2$. To avoid excessive repetitions, if the number of tokens in the nucleus in the previous step was less than $3$, we increase the temperature slightly.

Changing the condition values throughout generation allows dynamic conditioning. Although it does not form the main focus of this work, we informally experimented with changing the conditions smoothly and linearly, and provide a set of representative sound examples in the online supporting material, described in Section \ref{sec:eval}.

\section{Evaluation}
\label{sec:eval}
The evaluation of music generation models is a constantly evolving area of investigation and currently no consensus exists. In lieu of pursuing a subjective listening experiment which may be complex to replicate, we instead opt for objective, quantitative approaches to evaluation.

We evaluate our models using the metrics negative log-likelihood (NLL), top-1, and top-5 accuracies. While measuring top-$n$ accuracy, for each token, the model's output is considered accurate if the ground-truth class is within the top $n$ probabilities of the model's output. The evaluation configuration is the same as the training, namely using chunks with a length of $1216$, and calculating the loss for every single token in the target sequence. This is much more challenging than only predicting the next token given the full sequence, since, at the extreme, the model tries to predict the first note of a song, only given the \texttt{<START>} token. We ensure that the entirety of the test split is used by sequentially taking non-overlapping chunks, resulting in $1836$ chunks overall.

We additionally perform a quantitative evaluation on samples generated by our conditional models, by analyzing their emotional content, as done by Hung et al. \cite{emopia}. To this end, we first train a regression model to predict the emotion values of the samples from the training data split. The architecture of the regression model is a music transformer with $8$ layers, and the final layer outputs two continuous values, namely the valence and the arousal. Then using the trained conditional generation models, we perform inference using a collection of conditions, and later predict the emotional content of the generated samples using the trained regression model. As the error metric, we use the normalized $L_1$-distance between the predictions of the regression model and the conditions that were fed during inference. To make a fair comparison against the \textit{discrete-token} model, the condition values are chosen as the midpoints of the bins used by the discrete condition tokens, namely $-0.8$, $-0.4$, $0$, $0.4$, and $0.8$. Using a combination of $5$ values for valence, and $5$ values for arousal, we end up with a collection of $25$ condition value pairs. For each model and each condition, we generate $8$ samples without ``cherry-picking'' and report the average error. Each sample has $4096$ tokens. The regression model takes inputs with a length of $1216$, similar to the generator. Samples are fed into the regression model using a sliding window with $50\%$ overlap, and outputs are averaged. The overall scheme for evaluating generated samples is visualized in Figure \ref{fig:eval_inference}.

We also make the generated samples available online\footnote{\url{serkansulun.com/midi}}. As explained previously, using the conditional models, we generate $200$ samples for each. Since the generated melodies have a fixed number of tokens, their durations in time are inversely proportional to their tempo. The melodies generated with the lowest arousal conditions are on average $159.6$ seconds long. We also generate $200$ more samples using the vanilla model, with no conditioning. To demonstrate the dynamic conditioning ability of our models qualitatively, we additionally generate $4$ samples per model, using the \textit{continuous-token} and \textit{continuous-concatenated} models. Overall, we present $804$ samples. The midi files are rendered into mp3 format using the Fluidsynth software\footnote{\url{https://www.fluidsynth.org/}} and FluidR3\_GM soundfont\footnote{\url{https://archive.org/details/fluidr3-gm-gs}}. 

\Figure[t]()[width=0.99\linewidth]{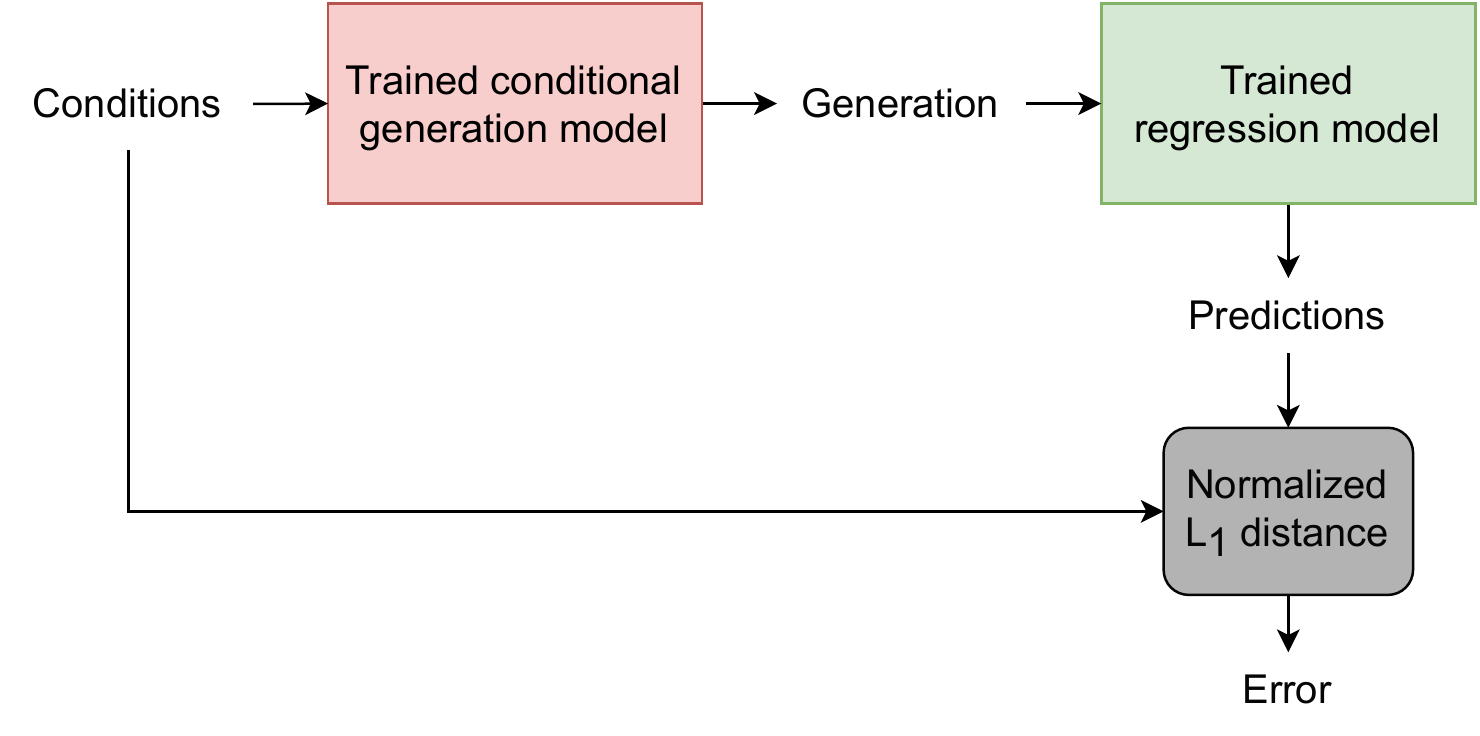}
{Evaluation of inference. \label{fig:eval_inference}}

\section{Results and Discussion}
\label{sec:results}
The performance of the models according to the prediction accuracy-based evaluation can be seen in Table \ref{table:eval}. The \textit{continuous-concatenated} model outperforms other models, including the state-of-the-art \textit{discrete-token} model, across all metrics, by a considerable margin especially for negative log-likelihood and top-1 accuracy. 

\begin{table}[h]
\caption{Performance of the models during evaluation. NLL refers to negative log-likelihood, where lower is better. Top-1 and Top-5 refer to the accuracy, where higher is better.}
\label{table:eval}
\centering
\begin{tabular}{|l|c|c|c|}
\hline
Model                    & NLL    & Top-1  & Top-5  \\  \hline\hline
vanilla                     & 0.7445 & 0.7784 & 0.9513 \\ \hline
discrete-token              & 0.7375 & 0.7885 & 0.9536 \\ \hline
continuous-token            & 0.7122 & 0.7895 & 0.9545 \\ \hline
continuous-concatenated     & \textbf{0.7075} & \textbf{0.7913} & \textbf{0.9548} \\ \hline
\end{tabular}
\end{table}

In Table \ref{table:infer} we show the results for the regression-based evaluation and demonstrate that \textit{continuous-concatenated} model outperforms others in terms of its ability to convey emotion. Note, here the vanilla approach is not included since it is not conditioned on any emotion information.

\begin{table}[h]
\caption{Performance of the conditional models during inference. Error refers to the normalized $L_1$ distance between conditions fed during inference and the output of the trained regression which consumes the generated samples.}
\label{table:infer}
\centering
\begin{tabular}{|l|c|}
\hline
Model             & Error           \\ \hline\hline
discrete-token    & 0.2164          \\ \hline
continuous-token  & 0.1951          \\ \hline
continuous-concatenated & \textbf{0.1948} \\ \hline
\end{tabular}
\end{table}

When considering the difference in performance among the presented models, 
we speculate that the main shortcoming of the discrete-token and continuous-token models, as opposed to the \textit{continuous-concatenated} model, is that they attribute the same importance to the condition values as the tokens in the sequence. We argue that while each token in the sequence is more useful in making local predictions, i.e., predicting the tokens that are near, the condition values have a global usage since they directly affect the entire generated sample. Our proposed \textit{continuous-concatenated} model can fully exploit the condition information by incorporating it into every single embedding of the input sequence for the transformer. Both our proposed methods can additionally use continuous-valued conditions, and thus allow much finer control over the generation process. These methods also permit the user to change the conditions throughout the generation, theoretically creating more complex and progressive compositions. 

Overall, we take a step towards establishing a more explicit connection between emotion and symbolic music. We identify the main challenge as the lack of symbolic music data paired with emotion labels. To tackle this, we augment the Lakh MIDI dataset, one of the largest symbolic music datasets available, with continuous-valued labels from the Spotify Developers API. While low-level features such as note density can loosely represent arousal, it is challenging to derive a similar representation for the valence dimension, especially using continuous-valued labels. We show that the valence values in our proposed dataset are indeed useful in bridging this gap, allowing us to generate long, coherent, multi-instrument symbolic music based on continuous-valued conditions taken arbitrarily from the valence-arousal plane.

For reproducibility and to help future research, we open-source our dataset and the code that we used to prepare it\footnote{\url{https://github.com/serkansulun/midi-emotion}}. In the NLP community, training large transformers from scratch is a rare practice that is typically replaced by transfer learning, namely by fine-tuning open-source pre-trained models. However, a similar phenomenon does not exist in the field of symbolic music generation. Thus, we additionally open-source our trained models to allow other researchers to cut down on the time and resources for training, with transfer learning. To the best of our knowledge, ours are the largest open-source symbolic music generation models, that are trained on the largest multi-instrument symbolic music dataset, in the literature.

In future work we intend to investigate the potential for conditional music generation directly in the audio domain. In this way we seek to build upon existing models such as WaveNet \cite{wavenet}, SampleRNN \cite{samplernn}, and Jukebox \cite{jukebox}.
One particular limitation in raw audio generation is the audio quality of the output, and the considerable computational demands of generating high resolution audio signals. On this basis, we envisage the potential for a two-stage approach which can combine lower quality raw audio generation with audio enhancement techniques, e.g., \cite{jukebox, kuleshov, sulun}, as well as exploring hybrid approaches which simultaneously leverage symbolic and audio data.

\section*{Acknowledgement}
The graphics card used for this research was donated by the
NVIDIA Corporation.

\bibliographystyle{ieeetr}
\bibliography{references.bib}

\begin{IEEEbiography}[{\includegraphics[width=1in,height=1.25in,clip,keepaspectratio]{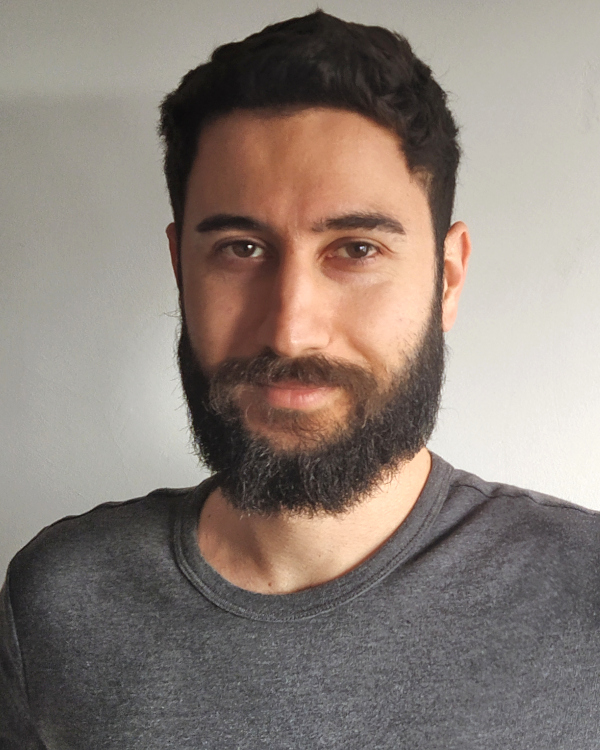}}]{\textbf{Serkan Sulun}} received the B.S. degree in electronics engineering from Sabanci University, Istanbul, Turkey, in 2014 and the M.S. degree in electrical and electronics engineering from Koc University, Istanbul, Turkey in 2018. He is currently pursuing the Ph.D. degree in electrical and computer engineering at the University of Porto, while working at Institute for Systems and Computer Engineering, Technology and Science (INESC TEC), in Porto, Portugal. His main research interest is multimedia signal processing using machine learning; specifically MIDI, audio, image, and video processing using deep neural networks.
\end{IEEEbiography}

\begin{IEEEbiography}[{\includegraphics[width=1in,height=1.25in,clip,keepaspectratio]{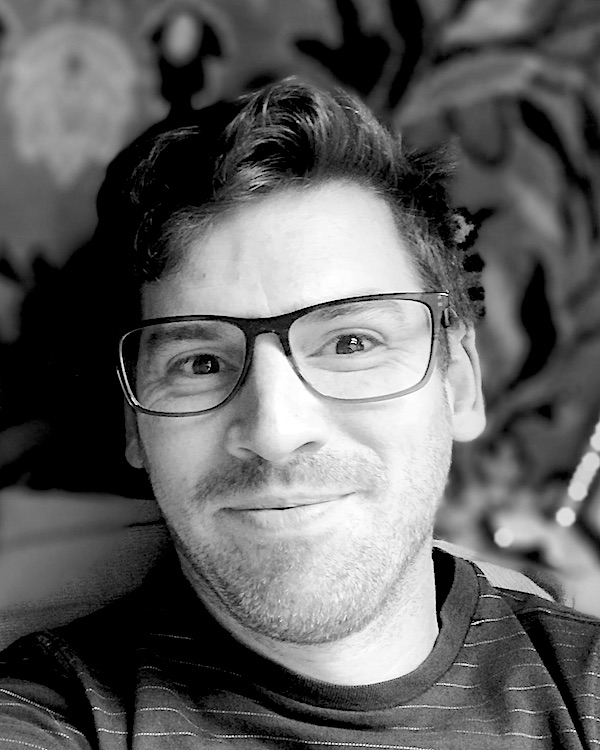}}]{\textbf{Matthew E. P. Davies}} received the B.Eng. degree in computer systems with electronics from King’s College London, U.K., in 2001 and the Ph.D. degree in electronic engineering from Queen Mary University of London, U.K., in 2007. From 2007 until 2011, he was a post-doctoral researcher in the Centre for Digital Music, QMUL. In 2013, he worked in the Media Interaction Group, National Institute of Advanced Industrial Science and Technology (AIST). From 2014-2019 he coordinated the Sound and Music Computing Group at INESC TEC, and is currently a researcher in the Centre for Informatics and Systems of the University of Coimbra (CISUC). His main research interests include music information retrieval, evaluation methodology, and creative music systems.
\end{IEEEbiography}

\begin{IEEEbiography}[{\includegraphics[width=1in,height=1.25in,clip,keepaspectratio]{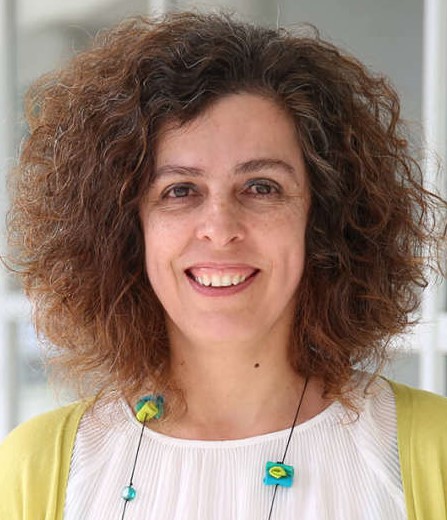}}]{\textbf{Paula Viana}}, (Senior Member, IEEE) received the Ph.D. degree in Electrical and Computer Engineering from University of Porto, in 2008. She is currently a Coordinator Professor with the School of Engineering, Polytechnic of Porto, and the Head of the Multimedia Communication Technologies at INESC TEC. She has over 30 years of experience in the area of multimedia content analysis and management, computer vision, and multimedia metadata. She has been coordinating the participation of INESC TEC in several national and European projects. She is the author of several publications, an active reviewer for journals, conferences and of European and Portuguese research projects. She has been involved in the organization of several scientific events, including the Immersive Media Experiences Workshop Series (2013–2015) at ACM Multimedia.
\end{IEEEbiography}

\EOD
\end{document}